\documentclass[
    ,final            
  ]
  {aipproc}

\layoutstyle{6x9}
\newcommand{\be}{\begin{equation}}

\newcommand{\ee}{\end{equation}}

\newcommand{\bea}{\begin{eqnarray}}
\newcommand{\eea}{\end{eqnarray}}

\begin{document}

\title{QCD, New Physics and Experiment}


\author{Giuseppe Nardulli}{
  address={Department of Physics, University of Bari and
INFN-Bari\\
Via E. Orabona 4, 70126, Bari, Italy\\
E-mail: giuseppe.nardulli@ba.infn.it} }

\begin{abstract}
I give a summary of Section E of the seventh edition of the
Conference {\it Quark confinement and the hadron spectrum}. Papers
were presented on different subjects, from spectroscopy, including
pentaquarks and hadron structure, to the quest for physics beyond
the standard model.
\end{abstract}

\maketitle

\section{Search for New Physics\label{sec:1}}
Before discussing how and where one should look for new physics, a
preliminary analysis of the status of the Standard Model (SM) is
needed. In Section E this was the task of T. Dorigo, who presented a
summary of SM tests at the Tevatron and P. Taras, who discussed the
BaBar experiment on the muon anomalous magnetic moment.

 As reported
by Dorigo, at present the top quark mass has been measured with an
accuracy of 1.2\%. The average result from Tevatron Run I/II is
$m_{top}=171.4\pm 2.1$ GeV. As a result of the increased accuracy on
$m_{top}$ the available parameter space for the Higgs mass within
the SM has been sharpened. The result of the Tevatron studies is
that the Higgs boson mass cannot be much larger than the present
limit of 114.4 GeV \cite{PDG}. The latest LEPEWWG results (summer
2006) are in fact $m_{higgs}=85^{+39}_{-28}$
 GeV, and $ m_{higgs}<166$ GeV at 95\% C.L.

Apart from indirect hints from radiative corrections, the Higgs
particle has been hunted in many different channels. Dorigo
presented some of these results. For example in $WH\to l\nu b b$
searches D0 extracts a limit for the cross section of 2.4 pb (at
95\%C.L.) for $m_{higgs}=115$ GeV with 380 pb$^{-1}$ of data; CDF
excludes cross sections above 3.4 pb with 1 fb$^{-1}$. These limits
are above SM cross sections and therefore  there is no exclusion
region in $m_{higgs}$ yet. To give another example, in D0 searches
for $H\to WW$ by selecting two high-$P_t$, isolated leptons
($ee,\,e\mu,\,\mu\mu$), with significant missing $E_t$, and little
jet activity, in $ee$ 11 events are seen (11.4 expected); in $
e\mu$: 18 seen (28.1 expected); in $\mu\mu$: 10 seen (10.5
expected). The expected SM Higgs signal is small and the limits are
dominated by systematic effects. Though statistics is not yet
sufficient to exclude definite $m_{higgs}$ regions, Tevatron Run II
preliminary data are getting closer to this result day by day. If
the Higgs boson exists and is light, it might be therefore
discovered within two or three years at Fermilab.

The results  presented by Taras were obtained by the BaBar
collaboration measuring the ratio ($\sqrt s$ is the c.m. energy)\be
R(s)\,=\,\frac{\sigma(e^+e^-\to{\rm
hadrons})}{\sigma(e^+e^-\to\mu^+\mu^-)}\ee using initial state
radiation, in the context of the study of the anomalous muon
magnetic moment $a_\mu$. It is defined by the relation $g-2=2a_\mu$
and comprises three components: the QED part $a_\mu^{QED}$, the
electroweak part $a_\mu^{EW}$, and the hadronic contribution
$a_\mu^{had}$. The components $a_\mu^{QED}$ and $a_\mu^{EW}$ are
computed with high precision (5 and 2 loops), so that the main
source of theoretical uncertainties is from $a_\mu^{had}$ and this
quantity is dominated by the integral\be\label{K}
a_\mu^{had,Lo}=\left(\frac{\alpha\,m_\mu}{3\pi}\right)^2
\int_{4m_\pi^2}^\infty\, ds\,K(s)\frac{R(s)}{s^2}\ee (apart from
$a_\mu^{had,Lo}$ there are other smaller contributions to
$a_\mu^{had,Lo}$ that tend to cancel out). In \eqref{K} $K(s)$ takes
values in the interval $(0.63,1)$ and $R(s)$ is obtained
experimentally summing up several channels, both non resonant
($2\pi$) and resonant ($\omega,\phi, J/\Psi$, etc.). The result
quoted by Paras is $a_\mu^{had,Lo}=(690.9\pm3.9_{exp}\pm
1.9_{rad}\pm 0.7_{QCD})\time 10^{-10}$. Existing experimental data
on $e^+e^-\to\pi^+\pi^-$ show some disagreement (KLOE data differ
appreciably from SND, CMD-2 data); moreover also measurements of
spectral functions obtained by $\tau$ decay
($\tau^-\to\pi^0\pi^-\nu_\tau$) and $e^+e^-$ show some discrepancy,
which must be eliminated. This makes the independent determination
with BaBar initial state radiation especially interesting. It must
be noted the BaBar offers a unique opportunity to get precision  of
2-4\% or even better than 1\% for the two charged pions mode (the
dominant mode, contributing by $\sim$ 73\% to $a_\mu^{had,Lo}$). If
one compares $a_\mu^{exp}$ obtained by SND, CMD-2 and BaBar, with
the result in the Standard Model $a_\mu^{SM}$: \bea
a_\mu^{exp}&=&(11659208.0\pm 6.3) \times 10^{-10}~,\cr
a_\mu^{SM}&=&(11659180.5\pm 5.6)\times 10^{-10}~,
 \eea
one finds a discrepancy of $3.3$
 standard deviations:\be a_\mu^{exp}-a_\mu^{SM}=(27.5\pm 8.4)
 \times 10^{-10}~. \ee This might be taken as a
 signal of new physics, but, on the basis of
 recent lattice QCD and $\tau$-based calculations,
 it is fair to see that the theoretical
 error has been probably underestimated
 (see Rubakov's talk in \cite{rubakov}).

 One of the
talks (J.Ulbricht) presented in Section E was devoted to tests of
non poinlike behavior of fermions. Measurements from various
experiments (LEP, TRISTAN, CDF, D0, UA2) have been used to search
for such a non point-like behaviour. In this way the group
responsible for this analysis (I. Dymnikova, U. Burch, J.Ulbricht,
C.H. LIN, S. Sakharov, J. Wu and J. Zhao) was able to put limits on
the energy scale $\Lambda$ of the direct contact interaction.
Ulbricht presented also models with excited fermions, contact
interaction and compositeness and he put constraints on the mass of
excited electron $m^*$, and on $\Lambda$. In more detail, excited
quarks with a mass between 80 and 570 GeV are excluded at 95 \%
confidence level. Using the UA2 data, according to his results one
can exclude excited $u^*$ and $d^*$ quarks with masses smaller than
288 GeV at 90\% CL. For EM interactions one gets limit on the mass
of a heavy electron: $m^*=308\pm 56$ GeV and for the finite size of
the electron a limit of $\Lambda=1253.2 \pm 226$ GeV, corresponding
to a size $r\approx 16\times 10^{-18}$cm . For  EW interaction the
most stringent limits for the quarks are $r_q<2.2\times 10^{-18}$cm,
for the leptons $r_l<0.9\times 10^{-18}$ cm, and the form factor
puts a limit on the electron size of $r_e<28 \times 10^{-18}$cm.
Finally a scheme to describe all fundamental particles as extended
objects of a finite geometrical size was presented.

Search for physics beyond SM is one of the missions of the
    future Large Hadron Collider at CERN. R. Mackeprang presented a
    talk on the quest for supersymmetry at Atlas. As a matter of
    fact this detector is sensitive to a broad spectrum of SUSY
    phenomenology, which strongly qualifies it for these studies.
    He presented some examples of ATLAS analyses with
    different scenarios and concluded that at 10 fb$^{-1}$ this
    experiment would be sensitive to a SUSY scale not larger than 2 TeV.
    Needless to say that also in this case a knowledge of the SM
    background as precise as possible will be of invaluable help in the
    identification of signals of SUSY, if they will be there.
    SUSY searches were mentioned also by H. Fox who presented a survey of
    results from the D0 experiment concerning new physics.
Besides results on supersymmetry he presented limits on the masses
of extra gauge bosons $W^\prime$ and $Z^\prime$. The results
obtained are $m_{W^\prime}>965$ GeV  and $m_{Z^\prime}>850$  GeV,
both at 95\% CL. He also presented results for physics beyond SM
with extra dimensions. Actually the analysis, inconclusive so far,
was limited to large extra dimensions studied in the context of
Russel-Sundrum models. Search for Extra Dimensions will be pursued
in the future with the ATLAS and CMS Detectors at the LHC , see e.g.
\cite{Shmatov}.

I wish to make here a digression on extra dimensions and QCD. The
linkage between the two is especially interesting for an
understanding of Quantum ChromoDynamics in the strong coupling
regime. Though the topic was not explicitly presented, it was part
of the discussion and deserves a mention. These developments are
related to the gauge/gravity correspondence \cite{maldacena}. In the
last few years it has been realized that such a correspondence can
be used to get information on QCD in the nonperturbative regime. In
particular by the term AdS/QCD one identifies the mapping between
D=4 strongly coupled gauge theories and gravitational theories in
D=5 with an  anti-de Sitter gravity background. This is a fast
growing field of study (for a review talk see \cite{marchesini}) in
which one can distinguish two different approaches. In the up-bottom
approach one starts from a string theory with an appropriate
background chosen so that some fundamental properties of QCD are
reproduced. By the bottom-up approach, starting from QCD one tries
to constrain the dual theory using the gauge/gravity correspondence.
To give just a few recent examples, these methods have been used to
shed light on the parameters of low energy effective theory of
Goldstone bosons \cite{Erlich}, linearity of Regge trajectories
\cite{Karch}, the heavy quark potential \cite{Andreev:2006ct}, the
thermal phase transition \cite{Andreev} or the BFKL Pomeron
\cite{Brower}. These few examples suffice to show the interest of
this approach and convince the reader that more results will be
certainly obtained in the near future.
\section{Hadron Structure, Heavy quark physics
and the pentaquark}Hadron structure is also a lively field of study;
more than 50 papers were presented at International Conference on
High Energy Physics at Moscow, August 2006.  The related arguments
discussed in Section E were various. In particular ample space was
given to structure functions and parton density functions from
neutral and charged currents and jets. The link to new physics stems
from the fact that parton distribution functions measured at the
existing accelerators are an essential piece of information for LHC.
These items were covered by Osipenko's and Cwiok's talks. Related
aspects are the nucleon spin structure (Livingston's talk) and
improvements by the inclusion of the Polyakov loop in the
description of low energy QCD by effective field theories (Megias).

Another topic discussed in the Section was the status of the
recently found charmonium-like states, i.e. the states X,Y and Z.
Let us start with the state X(3872). It was reviewed by S. Ricciardi
from BaBar and by A. Zupanc from the Belle collaboration. The
average mass of this state (also observed at CDF and D0) is
$3871.2~\pm~0.5$ MeV, with a width $\Gamma~<~2.3$ MeV. Its quantum
numbers were also established: $J^{PC}=~1^{++}$. This assignment
follows from the following considerations. First, since the decay
$X\to \gamma J/\Psi$ is observed, then the $X$ state must have
$C=+1$. Second, the decay $X\to \pi^+\pi^- J/\Psi$ is also observed.
The part of the $2\pi$ invariant mass spectrum that can be ascribed
to a $\rho^0$ decay is consistent with S-wave decay of the $X$
state. From this the assignment $P=+1$ follows. Finally the angular
distribution in this channel is incompatible with $J=0$ and
therefore the only remaining possibility is $J=1$ or $J=2$. However
if the peak in the $D^0\bar{D}^0\pi^0$ decay channel at 3875.4 MeV,
i.e. at only $2\sigma$ from the mass of $X(3872)$, is interpreted as
due to our state, the $J=2$ should be excluded and the only
remaining possibility is $J=1$.

Several interpretations of this state have been advanced in the
literature. One possibility is that it is a $\chi_{c1}^\prime$
state, but this is unlikely because of this result \be \frac{{\cal
B}(X\to \gamma J/\Psi)}{{\cal B}(X\to \pi^+\pi^-
J/\Psi)}=~0.19\pm0.07\label{average} ~,\ee which is an average of
BaBar and Belle results and is too small to be compatible with this
identification. Another possibility is that this state comprises
four quarks, more exactly two diquarks \cite{Maiani:2004vq}. If this
interpretation is correct, then the existence of additional states
can be predicted. At present these new states have not been seen.
Moreover the mass difference between the two neutral states is
larger by two $\sigma$ than experimental data \cite{Swanson:2006st}.
While it is too early to get definite conclusions, it is fair to say
that a more economical interpretation is that in terms of a
molecular $D^0\bar D^{*0}$ state
\cite{tornqvist,Close:2003sg,Wong:2003xk,Pakvasa:2003ea}.

Let us now consider the state Y(3940), observed by Belle in the
decay mode $B~\to~ K~\omega~ J/\Psi$. Its reported mass is
$M=3943\pm11\pm13$ MeV, with a width $\Gamma= 87\pm22\pm26$ MeV. Its
interpretation as a charmonium $c\bar c$ state is possible, but
should be corroborated by the observation of the decay mode $Y\to
D^{(*)}\bar D^{(*)}$, which has not yet been seen (on the contrary
the decay mode $Y\to\omega J/\Psi$ has a large branching ratio). It
could be a $c\bar c$-gluon hybrid since in this case the decay mode
$Y\to D^{(*)}\bar D^{(*)}$ would be suppressed, but the difficulty
is in the predicted mass of such a state, around 4.3-4.5 GeV from
lattice QCD computations \cite{Bernard:1997ib,Mei:2002ip},
significantly larger than the measured value. One can therefore
conclude that more data are needed before an identification of this
state as a $c\bar c g$ can be made. Zupanc also discussed the new
state $X(3940)$ found in double charm production, probably different
from $Y(3940)$ and the state $Z(3930)$, whose possible
interpretation is $\chi_{c2}^\prime$. I refer the interested reader
to his talk, as reported in these proceedings, for a detailed
discussion of these results.

S. Ricciardi discussed other two charmonium-like states. The first
one is the state $Y(4260)$ observed by BaBar in $e^+e^-\to
(\gamma)Y(4260)\to~J/\Psi\pi^+\pi^-$ (with no $\gamma$ detection).
BaBar finds for this state $M=4259\pm 8^{+2}_{-6}$ MeV and
$\Gamma=88\pm 23^{+6}_{-4}$ MeV. This state is confirmed by Cleo-III
and Belle (the mass measured by Belle is however 2.5$\sigma$, i.e.
36 MeV, higher). Data show a large coupling to $J/\Psi\pi\pi$, which
renders puzzling its interpretation as charmonium-like. Other
interpretations, in terms of  a tetraquark state
\cite{Maiani:2005pe} or a hybrid, should be seriously considered.
The other structure is seen in $e^+e^-\to
(\gamma)\psi(2S)\pi^+\pi^-$ at a mass $4.35$ GeV. Data are not
sufficient to draw conclusions about the nature of the structure or
its consistency with the previously observed Y(4260).

Among the various topics discussed by Ricciardi, let me mention the
problem of the state $D^*_s(2860)$. This charmed meson, seen through
a fit to DK mass spectra, has mass and width as follows: $M = 2856.6
\pm 1.5 \pm 5.0$ MeV, and  $\Gamma = 47 \pm 7 \pm 10$ MeV. Since it
decays into two pseudoscalar mesons, its spin-parity assignment can
be $J^P=0^+,~1^-,~2^+~,3^-$. At present the decay mode $D^*K$ has
not been observed and this means, according to the authors in
\cite{Colangelo:2006rq}, that the assignments $1^-,~2^+$ are not
favored. Between the remaining alternatives $0^+,~3^-$ the latter
seems more likely because in the former case the decays would occur
in $S-$wave and the resonance would be broad. On the other hand the
existence of a narrow state with $\Gamma=35-140$ MeV, Strangeness=0
and  $ M $ around 2.8 GeV was predicted already in 2000
\cite{Colangelo:2000jq}. Its strange partner would have therefore a
mass compatible with the new state. The calculation in
\cite{Colangelo:2000jq} was based on QCD sum rules in the framework
of the Heavy Quark Effective Theory (HQET) as applied to mesons
comprising one heavy quark \cite{Casalbuoni:1996pg}. HQET classifies
these states according to $s_\ell^P$, the total angular momentum of
the light degrees of freedom. For $s_\ell^P = 3/2^-$ one has $J^P =
1^-,~ J^P = 2^-$; there are other two narrow states with $s_\ell^P =
5/2^- $ and therefore with $ J^P = 2^- ,~ J^P = 3^-$. For a $3^-$
state, the decay width into $DK$ is predicted to be small because it
occurs as $ F-$wave; on the other hand missing observation of the
$2^-$ partner might be explained by its likely mixing with the broad
state. Other talks on heavy quarks were on further results from
BELLE, in particular on the $Y(4260)$ state (G.Pakhlova).

Another topic discussed in Section E was the negative result from
CLAS in the search for pentaquark with increased statistics (R.
Gothe's talk). Pentaquarks were one of the highlights of the Section
E of the 2004 Conference on Quark Confinement and the Hadron
Spectrum \cite{Nardulli:2004gz}. In photoproduction experiment off
proton no significant signal for  $\Theta^{+}$ or $\Theta^{++}$ is
seen by CLAS. Also for photoproduction off deuteron: $\gamma d\to
\Theta^+K^-$, followed by $\Theta^+\to nK^+$, CLAS has now negative
results, differently from their previously released analysis. In
particular previous peak could not be reproduced under similar
circumstances. Therefore the statistical significance of the old
peak is reduced from 5.2$\sigma$ to 3.1$\sigma$ when new data are
used as background. While the existence of the pentaquark cannot be
excluded, stringent upper limits on total and differential cross
sections were set. In particular, present  data, together with
phenomenological models, put an upper limit on the cross section for
$\gamma n\to\Theta^+K^-$ of around 3 nb (at 95\% CL). Let me note
that pentaquark searches were reported elsewhere at this Conference,
with mixed results from other experiments, see e.g. K. Daum's talk
given in Section B.

In conclusion, waiting for LHC, still a lot of physics can be done
and is actually done. In Section E reports were presented on several
active and promising areas: Heavy quarks, supersymmetry,
compositeness, excited states and the role of extra dimensions. As
to string physics, the new developments on AdS/QCD, if correct,
would represent a further step in the long and complicated journey
from strong interactions to gravity and back.

\begin{theacknowledgments}I thank P. Colangelo for most useful
comments on the spectroscopy of charmed states,
\end{theacknowledgments}


\begin{thebibliography}{99}
\bibitem{PDG}W. -M. Yao, et al.,``Review of Particle Physics",
 \emph{Journal\ of\ Physics\ G} \textbf{33}, 1 (2006).
\bibitem{rubakov}V. Rubakov, ``Conclusions and Outlook",
summary talk at the XXXIII International Conference on High Energy
Physics, July 26 - August 2, 2006, Moscow, Russia.
\bibitem{Shmatov}S. Shmatov, ``Search for Extra Dimensions with
ATLAS and CMS Detectors at the LHC", talk presented at the XXXIII
International Conference on High Energy Physics, July 26 - August 2,
2006, Moscow, Russia.
\bibitem{maldacena}J. M. Maldacena,  \emph{ Adv.\ Theor.\
Math.\ Phys.} \textbf{2},
  231 (1998).
  \bibitem{marchesini}G. Marchesini, ``QCD review", talk presented at the XXXIII
International Conference on High Energy Physics, July 26 - August 2,
2006, Moscow, Russia. 
    \bibitem{Erlich}J. Erlich. E. Katz, D. T. Son, and M. A.
  Stephanov, \emph{ Phys.\ Rev.\
Lett.} \textbf{95},
  261602 (2005).
\bibitem{Karch}A. Karch, E. Katz, D. T. Son, and M. A.
  Stephanov, \emph{ Phys.\ Rev.\
D} \textbf{74}, 015005 (2006) [arXiv:hep-ph/0602229].
\bibitem{Andreev:2006ct}
  O.~Andreev, and V.~I.~Zakharov,
  \emph{Phys.\ Rev.\ D}  \textbf{74}, 025023 (2006)
  [arXiv:hep-ph/0604204].
  \bibitem{Andreev}O. Andreev, and V. I. Zacharov, arXiv:hep-ph/0607026.
  \bibitem{Brower}R. C. Brower, J. Polchinski, M. J. Strassler, and
  C.-I Tan, arXiv:hep-th/0603115.
  \bibitem{Maiani:2004vq}
  L.~Maiani, F.~Piccinini, A.~D.~Polosa, and V.~Riquer,
  \emph{  Phys.\ Rev.\ D} \textbf{71},
  014028 (2005)  [arXiv:hep-ph/0412098].
\bibitem{Swanson:2006st}
  E.~S.~Swanson,
  \emph{Phys.\ Rept.\ } {\bf 429}, 243 (2006)
  [arXiv:hep-ph/0601110].
\bibitem{tornqvist}N. A. T\"ornqvist, arXiv:hep-ph/0308277.
\bibitem{Close:2003sg}
  F.~E.~Close, and P.~R.~Page,
  \emph{Phys.\ Lett.\ B} {\bf 578}, 119 (2004)
  [arXiv:hep-ph/0309253].
\bibitem{Wong:2003xk}
  C.~Y.~Wong,
  \emph{Phys.\ Rev.\ C } {\bf 69}, 055202 (2004)
  [arXiv:hep-ph/0311088].
\bibitem{Pakvasa:2003ea}
  S.~Pakvasa, and M.~Suzuki,
  \emph{Phys.\ Lett.\ B}  {\bf 579}, 67 (2004)
  [arXiv:hep-ph/0309294].
\bibitem{Bernard:1997ib}
  C.~W.~Bernard, {\it et al.}  [MILC Collaboration],
  \emph{Phys.\ Rev.\ D } {\bf 56}, 7039 (1997)
  [arXiv:hep-lat/9707008].
\bibitem{Mei:2002ip}
  Z.~H.~Mei, and X.~Q.~Luo,
     \emph{Int.\ J.\ Mod.\ Phys.\ A} {\bf 18}, 5713 (2003)
  [arXiv:hep-lat/0206012].
\bibitem{Maiani:2005pe}
  L.~Maiani, V.~Riquer, F.~Piccinini, and A.~D.~Polosa,
  \emph{Phys.\ Rev.\ D} {\bf 72}, 031502 (2005)
  [arXiv:hep-ph/0507062].
\bibitem{Colangelo:2006rq}
  P.~Colangelo, F.~De Fazio, and S.~Nicotri,
  \emph{Phys.\ Lett.\ B} {\bf 642}, 48 (2006)
  [arXiv:hep-ph/0607245].
\bibitem{Colangelo:2000jq}
  P.~Colangelo, F.~De Fazio, and G.~Nardulli,
  \emph{Phys.\ Lett.\ B} {\bf 478}, 408 (2000)
  [arXiv:hep-ph/0001200].
\bibitem{Casalbuoni:1996pg}
  R.~Casalbuoni, A.~Deandrea, N.~Di Bartolomeo, R.~Gatto, F.~Feruglio, and G.~Nardulli,
  \emph{Phys.\ Rept. }  {\bf 281}, 145 (1997)
  [arXiv:hep-ph/9605342].
\bibitem{Nardulli:2004gz}
  G.~Nardulli, ``QCD, hadrons and beyond'', edited by U. D'Alesio et
  al.,
  AIP Conference  Proceedings  756, American Institute of Physics,
  New York,
  2005, pp. 228-235 (2005)
  [arXiv:hep-ph/0411294].
   \end{thebibliography}
\end{document}